\documentclass[runningheads]{svmult}

\usepackage{graphicx}  
\usepackage{subeqnar}  
\usepackage{multicol}  
\usepackage{physprbb}  

\newcommand {\be} {\begin{equation}}
\newcommand {\ba} {\begin{eqnarray}}
\newcommand {\ee} {\end{equation}}
\newcommand {\ea} {\end{eqnarray}}

\begin{document}
\title*{Proton Structure Corrections
to Hydrogen Hyperfine Splitting}
\toctitle{Proton Structure Corrections
to Hydrogen Hyperfine Splitting}
%
%
\titlerunning{Proton Structure Corrections to Hhfs}
%
\author{Carl E. Carlson}
\authorrunning{Carl E. Carlson}
%
%
\institute{Physics Department, The College of William and Mary, \protect\newline
Williamsburg, VA 23187-8795, U.S.A.
}

\maketitle              

\begin{abstract}
The largest uncertainty in calculations of hydrogen ground-state hyperfine splitting comes from corrections due to proton stucture.  We review these corrections, with special mention of the inelastic, or polarizability, corrections which have been recently re-evaluated.  Summing up the arguably best current values for the calculated corrections leaves us 1--2 ppm short of the experimental data.  We speculate how one may explain this shortfall, and along the way broadly outline the derivations of the relevant formulas, attempting to explain how certain terms come to appear and commenting on the use of unsubtracted dispersion relations.
\end{abstract}

\section{Introduction}

Hyperfine splitting (hfs)
in the hydrogen ground state is measured
to 13 significant figures in frequency units~\cite{Karshenboim:2005iy},
    \be
    E_{\rm hfs}(e^-p) = {\rm 1\ 420.405\ 751\ 766\ 7  (9)\ MHz} \,.
    \ee
Theory is far from this level of accuracy, and theorists are hopeful of obtaining calculations 
accurate to a part per million (ppm) or so.   We are close to reaching this goal, but some improvement is still needed and there currently seems to be a few ppm discrepancy between the best calculations and the data.

The main uncertainty in calculating the hfs in hydrogen comes from the hadronic, or proton structure, corrections.  One can contrast this to the case of muonium, where the ``nucleus'' is a point particle, so that calculations are almost purely QED, and agreement between theory and experiment is about 0.1 ppm~\cite{Karshenboim:2005iy}.  

For ordinary hydrogen, as we have said, one must consider the proton structure, and find that it contributes about 40 ppm to the hfs.  Working out these contributions theoretically requires knowing details about proton structure that cannot currently be obtained from {\it ab initio} calculation.   Instead, one has to measure information about proton structure in other experiments, particularly experiments on elastic and inelastic electron-proton scattering.   Then calculations are done to relate the scattering information to the bound state energy. Recent new results have been driven by improvement in the data, including both new data for polarized inelastic scattering in kinematic regions of interest to hfs calculators and new analyses of the elastic scattering data. These will be outlined below.  

Historically, the elastic and inelastic contributions, the latter also called polarizability corrections, have often been treated separately, with the elastic corrections further divided into a nonrelativistic Zemach term and relativistic recoil corrections.   From a modern viewpoint, the elastic and inelastic corrections should be treated as a unit since the sum lacks certain ambiguities that exist in the individual terms.   The present discussion will focus on the polarizability contributions, but following the last remark, discussion of the Zemach and recoil corrections will not be omitted.

\section{Hyperfine splitting calculations}

The calculated hyperfine splitting in hydrogen is~\cite{Karshenboim:2005iy,Volotka:2004zu,dupays}
\ba
    E_{\rm hfs}(e^-p) =
    \big (1+\Delta_{\rm QED} + \Delta_{\rm weak}^p+\Delta_{\rm Str} \big)
    \, E_F^p \,,
\ea
where the Fermi energy is
\be
    E_F^p=\frac{8 \alpha^3 m_r^3 }{3\pi} \mu_B\mu_p
    	=	\frac{8  \alpha^4 m_r^3}{3 m_e m_p} \left( 1+\kappa_p \right) 	\,,	
\ee
with $m_r = m_e m_p / (m_p + m_e)$ being the reduced mass (and there are hadronic and muonic vacuum polarization terms~\cite{Volotka:2004zu} which are included as higher order corrections to the Zemach term below).  The QED terms are accurately calculated and well known.  They will not be discussed, except to mention that could be obtained without calculation.  The QED corrections are the same as for muonium, so it is possible to obtain them to an accuracy more than adequate for the present purpose using muonium hfs data and a judicious subtraction~\cite{Brodsky:2004ck,Friar:2005rs}. The weak interaction corrections~\cite{Eides:1995sq} also will not be discussed, and are in any case quite small.  We will discuss the proton structure dependent corrections,
\ba
\Delta_{\rm Str} = \Delta_{el} + \Delta_{inel} = \Delta_Z +  \Delta_R^p +  \Delta_{\rm pol} \,,
\ea
where the terms on the right-hand-side are the Zemach, recoil, and polarizability corrections.

\begin{figure}[htbp]
\begin{center}
\includegraphics[height=22mm]{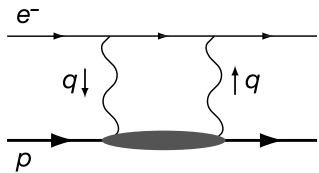}
\caption{Generic two-photon exchange diagrams, giving proton-structure corrections to hyperfine splitting.}
\label{twophoton}
\end{center}
\end{figure}

Generically, the proton structure corrections come from two-photon exchange, as diagramed in Fig.~\ref{twophoton}.  The diagram can be seen as Compton scattering of off-shell photons from an electron knit together with similar Compton scattering from the proton.  (We are neglecting the characteristic momentum of the bound electron.  This allows a noticeably simpler two-photon calculation than for a scattering process~\cite{Blunden:2003sp}.  One can show that keeping the characteristic momentum would give corrections of ${\cal O}(\alpha m_e/m_p)$ smaller than terms that are kept~\cite{IddingsP}.)


\subsection{Elastic terms: the Zemach correction}



In this author's opinion, the best calculation of the box diagram uses dispersion relations, even though there are questions about using dispersion relations in their unsubtracted form.  In a dispersive calculation, it is easy to consider the elastic and inelastic intermediate states simultaneously.  This one should do because the full calculation is well defined, even though historically terms have been shuttled between the ``elastic'' and ``inelastic'' contributions.  

At the outset, however, we will present results from a direct calculation of the elastic contributions, without dispersion theory.  The results have been obtained by a number of authors~\cite{Bodwin:1987mj,Martynenko:2004bt}, and follow after assuming a certain photon-proton-proton vertex which is plausible but which cannot be defended perfectly.  

\begin{figure}[htbp]
\begin{center}
\includegraphics[height=22mm]{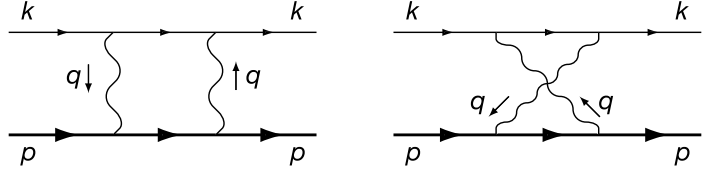}
\caption{Two-photon exchange diagrams for the elastic proton-structure corrections to hyperfine splitting.}
\label{boxes}
\end{center}
\end{figure}

The ``elastic'  contibutions are those where the hadronic intermediate state, the blob in Fig.~\ref{twophoton}, is just a proton.  The diagram specializes to Fig.~\ref{boxes}.  The photon-electron vertex is known, and we use~\cite{Bodwin:1987mj,Martynenko:2004bt}
\ba
\label{vertex}
\Gamma_\mu = \gamma_\mu F_1(q^2) + \frac{i}{2m_p} \sigma_{\mu\nu} q^\nu F_2(q^2)
\ea
for the photon-proton vertex with incoming photon momentum $q$.  Functions $F_1$ and $F_2$ are the Dirac and Pauli form factors of the proton, which are measured in elastic electron-proton scattering.  The normalization is $F_1(0) =1$ and $F_2(0) = \kappa_p$, where $\kappa_p$ is the proton's anomalous magnetic moment measured in proton magnetons.

The above photon-proton vertex is complete and correct only if the protons entering and exiting the vertex have physical, on-shell, momenta.  In a loop diagram, the intermediate proton is on-shell only at special values of momenta out of a continuum of momenta.  Hence one may feel some hesitation in using the results that follow.  However, there is also the (coming) dispersive calculation which only needs the vertices when all protons are on-shell, and so (again modulo questions surrounding dispersion relations with no subtraction) gives a reliable result.  Any terms in the ``elastic'' calculations that appear to require modification can be fixed by adding or subtracting terms in other parts of the quoted result.

The elastic contributions are separated as
\ba
\left.   \frac{E_{2\gamma}}{E_F} \right|_{el} = \Delta_Z + \Delta_R^p \,.
\ea
The separation is into non-relativistic and relativistic terms---``Zemach'' and ``recoil.''  Non-relativistic means the limit $m_p \to \infty$ with $m_e$ held fixed and with proton size held fixed; proton size information is embedded in the form factors $F_1$ and $F_2$.

The Zemach correction was worked out by Zemach in 1956~\cite{Zemach}; in modern form it is
\ba
\Delta_Z = \frac{8 \alpha m_r}{\pi} \int_0^\infty \frac{dQ}{Q^2} \ 
	\left[   G_E(-Q^2)  \frac{ G_M(-Q^2) }{ 1+ \kappa_p } - 1   \right]
		=  - 2 \alpha m_r  r_Z  \,,
\ea
where the last equality defines the Zemach radius $r_Z$ and we have used $Q^2=-q^2$.  The charge and magnetic form factors are linear combinations of $F_1$ and $F_2$,
\ba
G_M &=& F_1 + F_2  \,, \nonumber \\
G_E &=& F_1 - \frac{Q^2}{4 m_p^2}  F_2  \,.
\ea

Table~\ref{table:one} gives the evaluated Zemach radius $r_Z$ and correction $\Delta_Z$ for two modern form factor fits, and for the dipole fit, which is out of date and included only because it is a common benchmark.  We believe the Ingo Sick fit is best for the purpose at hand, because the Zemach integrals depend mainly on the form factors at low $Q^2$, and Sick's fit concentrates on the low $Q^2$ scattering data.  


\begin{table}[htdp]
\caption{Values of the Zemach radius and the Zemach corrections for selected fits to proton elastic form factors. (The Zemach term $\Delta_Z$ includes a 1.53\% correction from higher order electronic contributions~\cite{Karshenboim:1996ew}, as well as a +0.07 ppm correction from muonic vacuum polarization and a +0.01 ppm correction from hadronic vacuum polarization~\cite{Volotka:2004zu}.)}
\begin{center}

\begin{tabular}{lll}      
Form factor  \qquad \qquad &    $\  r_Z$ (fm) \qquad  \qquad & \ \  $\Delta_Z$ (ppm)  \\
\hline
Kelly~\cite{sickandkelly}  &  1.069(13)   &  $-40.93(49)$   \\
Sick~\cite{sickandkelly}   &  1.086(12)  &  $-41.59(46)$  \\
dipole                         &  1.025      &  $-39.24$   \\
\hline
\end{tabular}

\end{center}
\label{table:one}
\end{table}%



\subsection{Elastic terms: recoil corrections}


The relativistic elastic corrections $\Delta_R^p$ are known as recoil corrections.  
They depend on the form factors and hence are part of the proton structure corrections.  However, evaluating $\Delta_R^p$ with different form factor representations based on fits to the scattering data reveals that its numerical value is fairly stable (to about $\pm 0.15$ ppm) by present standards.

The full result is
\ba
\label{eq:recoil}
\Delta_R^p &=&  \frac{ \alpha m_e m_p} { 2(1+\kappa_p) \pi (m_p^2-m_e^2) } \ \times
					\nonumber	\\[1.25ex]
&& \times \ \ 
	\Bigg\{
	\int_0^\infty    \frac{ d Q^2 }{ Q^2 }
	\left(     \frac{\beta_1(\tau_p)-4\sqrt{\tau_p}}{\tau_p} 
		- \frac{\beta_1(\tau_e)-4\sqrt{\tau_e}}{\tau_e}		\right)
	F_1(-Q^2) G_M(-Q^2)  
					\nonumber	\\[1.25ex]
	&&	\hskip 8mm
	+ \ 3 \int_0^\infty  \frac{ d Q^2 }{ Q^2 } \   \Big( \beta_2(\tau_p) - \beta_2(\tau_e)	\Big) \ 
	F_2(-Q^2) G_M(-Q^2)	\Bigg\}
					\nonumber	\\[1.25ex]
&& + \ \ 
	\frac{2 \alpha m_r}{\pi m_p^2} 
	\int_0^\infty dQ \,
	F_2(-Q^2) \frac{G_M(-Q^2)}{1+\kappa_p}
					\nonumber	\\[1.25ex]
&& - \ \ \frac{ \alpha m_e } { 2(1+\kappa_p) \pi m_p }
	\int_0^\infty	 \frac{ d Q^2 }{ Q^2 } \ \beta_1(\tau_e) F_2^2(-Q^2)		\,,
\ea
where $\beta_{1,2}$ are auxiliary functions that were first found useful in discussing the inelastic terms~\cite{Iddings,Drell:1966kk,DeRafael:mc,Gnaedig:qt,Faustov:yp,Nazaryan:2005zc},
\begin{eqnarray}
\beta_1(\tau) &=& -3\tau + 2\tau^2
    + 2(2-\tau)\sqrt{\tau(\tau+1)}  = \frac{9}{4} \beta(\tau)	\,,
                    \nonumber \\
\beta_2(\tau) &=& 1 + 2\tau - 2 \sqrt{\tau(\tau+1)}		\,.
\end{eqnarray}
These are used with the notation,
\be
\tau_p \equiv \frac{Q^2}{4m_p^2} \,,	\quad
\tau_e \equiv \frac{Q^2}{4m_e^2} \,.	
\ee

The reasons for showing the whole formula for the recoil corrections is partly to show it is not so long (it sometimes appears more forbidding, {\it Cf.}~\cite{Bodwin:1987mj}), to explicitly see the form factor dependence, and to display the $F_2^2$ term in the last line.  The $F_2^2$ term is noteworthy because it is absent in a dispersive calculation, in contrast to all the other terms, which come forth unchanged.

All the integrals are finite, although some require the form factors to prevent ultraviolet divergence,  and some would be infrared divergent if $m_e \to 0$.  Of minor interest, the penultimate line could be subsumed into the Zemach correction, if the Zemach correction were to be written in terms of $F_1 G_M$ rather than $G_E G_M$.  


\subsection{Inelastic terms: polarizability corrections}


When the blob in Fig.~\ref{twophoton} is not a lone proton,  we obtain inelastic contributions or polarizability contributions~\cite{Iddings,Drell:1966kk,DeRafael:mc,Gnaedig:qt,Faustov:yp,Nazaryan:2005zc}. The inelastic contributions are not calculable {\it ab initio}.  Instead, one relates them to the amplitude for forward Compton scattering of off-shell photons off protons,  given in terms of the matrix element
\begin{eqnarray}			
T_{\mu\nu}(q,p,S) = \frac{i}{2\pi m_p} \int d^4\xi \ e^{iq{\cdot}\xi}
	\left\langle pS \right| j_\mu(\xi) j_\nu(0) \left| pS \right\rangle \,,
\end{eqnarray}
where $j_\mu$ is the electromagnetic current and the states are proton states of momentum $p$ and spin 4-vector $S$.  The spin dependence is in the antisymmetric part
\begin{eqnarray}			
\label{eqn:ta}
T^A_{\mu\nu} = \frac{i}{m_p\nu} \, \epsilon_{\mu\nu\alpha\beta} q^\alpha 
	\left[ \left(H_1(\nu,q^2) + H_2(\nu,q^2) \right) S^\beta - H_2(\nu,q^2) 
		\frac{ S{\cdot}q \ p^\beta }{ p{\cdot}q }		\right]  \,.
\end{eqnarray}
There are two structure functions $H_1$ and $H_2$ which depend on $q^2$ and on the photon energy $\nu$, defined in the lab frame so that $m_p \nu = p\cdot q$.

There is an optical theorem that relates the imaginary part of the forward Compton amplitude to the cross section for inelastic scattering of off-shell photons from protons.  The relations precisely are
\ba			
{\rm Im\,} H_1(\nu,q^2) = \frac{1}{\nu} \  g_1(\nu,q^2)
			\qquad  {\rm and} \qquad 
{\rm Im\,} H_2(\nu,q^2)  = \frac{m_p}{\nu^2} \   g_2(\nu,q^2)   \,,
\ea
where $g_1$ and $g_2$ are functions appearing in the cross section and are measured~\cite{Anthony:2000fn,Anthony:2002hy,Fatemi:2003yh,models,deur}  at SLAC, HERMES, JLab, and elsewhere.

Using the Compton amplitude in terms of $H_1$ and $H_2$, Eq.~(\ref{eqn:ta}), in evaluating the inelastic part  of the two-photon loops gives
\ba
\Delta_{\rm pol} &=& \left. \frac{E_{2\gamma}}{E_F}  \right|_{inel}
	= \frac{2\alpha m_e}{ ( 1+\kappa_p) \pi^3 m_p  }
		\int \frac{d^4Q}{(Q^4+4m_e^2 Q_0^2) Q^2} \ \times 
				\\[1.25ex]  \nonumber
&& \hskip 10mm  \times \ 
\Big\{ (2Q^2 + Q_0^2) H^{inel}_1(iQ_0,-Q^2) - 3 Q^2 Q_0^2 H^{inel}_2(iQ_0,-Q^2) \Big\} \,,
\ea
where we have Wick rotated the integral so that $Q_0 = -i \nu$, $\vec Q = \vec q$, and $Q^2 \equiv Q_0^2 + \vec Q^2$.
Since $H_{1,2}$ are not measured, we obtain them from a dispersion relation, which will discussed in a subsequent section.  Assuming no subtraction,  
\ba
H^{inel}_1(\nu_1,q^2) = \frac{1}{\pi} \int_{\nu_{th}^2}^\infty d\nu 
	\frac{ {\rm Im\,} H_1(\nu,q^2) } { \nu^2 -\nu_1^2 }  \,,
\ea
where the integral is only over the inelastic region 
($\nu_{th}=m_\pi + (m_\pi^2+Q^2)/(2m_p)$), and similarly for $H_2$.

Putting things together, neglecting $m_e$ inside the integral, and  integrating what can be integrated, one obtains the expression
\ba
    \Delta_{\rm pol}=\frac{ \alpha m_e}{2 (1+ \kappa_p) \pi m_p}
    (\Delta_1+\Delta_2),
\ea
where, with $\tau = \nu^2/Q^2$,			
\ba
  \Delta_1 &=& \frac{9}{4}\int_0^\infty \frac{dQ^2}{Q^2}\left\{F_2^2(-Q^2) +4 m_p
	\int_{\nu_{th}}^\infty	 \frac{d\nu}{\nu^2} \beta(\tau)  g_1(\nu, -Q^2)
		\right\}   \,,
					\\[1ex]	\nonumber
\Delta_2 &=& -12m_p  \int_0^\infty \frac{dQ^2}{Q^2}
	\int_{\nu_{th}}^\infty	 \frac{d\nu}{\nu^2} \beta_2(\tau)  g_2(\nu, -Q^2) .
\ea

The integral for $\Delta_1$ is touchy.  Only the second term comes from the procedure just outlined. The first arises when one applies the dispersive calculation also to the elastic corrections, and discovers the $F_2^2$ term pointed out earlier in Eq.~(\ref{eq:recoil}) is absent.  It was then thought convenient to add the first term as seen above, and then subtract the same term from the recoil contributions.  This leaves the elastic corrections exactly as already shown.  This stratagem also allows the electron mass to be taken to zero in $\Delta_1$.  The individual $dQ^2$ integrals in $\Delta_1$ diverge (they would not had the electron mass been kept), but the whole is finite because of the Gerasimov-Drell-Hearn (GDH) sum rule~\cite{Gerasimov:1965et,Drell:1966jv}, 
\ba		
4 m_p  \int_{\nu_{th}}^\infty   \frac{d\nu}{\nu^2}  \, g_1(\nu, 0) = - \kappa_p^2  \,,
\ea
coupled with the observation that the auxiliary function $\beta(\tau)$ becomes unity as we approach the real photon point.

The polarizability expressions have some history. A short version is that considerations of $\Delta_{\rm pol}$ were begun by Iddings in 1965~\cite{Iddings}, improved by Drell and Sullivan in 1966\cite{Drell:1966kk}, and given in present notation by de Rafael in 1971~\cite{DeRafael:mc}.  But no sufficient spin-dependent data existed, so it was several decades before the formula could be evaluated to a result incompatible with zero.  In 2002, Faustov and Martynenko became the first to use $g_{1,2}$ data to obtain results inconsistent with zero~\cite{Faustov:yp}.  Their 2002 result was
\ba		
\Delta_{\rm pol}  = (1.4 \pm 0.6) {\rm\  ppm}
\ea
However, they only used SLAC data and $\Delta_1$ and $\Delta_2$  are sensitive to the behavior of the structure functions at low $Q^2$.   Also in 2002 there appeared analytic expressions for $g_{1,2}$ fit to data by Simula, Osipenko, Ricco, and Taiuti~\cite{Simula:2001iy}, which included JLab as well as SLAC data.  They did not at that time integrate their results to obtain $\Delta_{\rm pol}$.  Had they done so, they would have obtained 
$\Delta_{\rm pol}  = (0.4 \pm 0.6) {\rm\  ppm}$~\cite{Nazaryan:2005zc}.

We now have enough information to discover a bit of trouble. Table~\ref{table:two} summarizes how things stood before the 2005/2006 re-evalations of $\Delta_{\rm pol}$.  The sum of all corrections is $1.59 \pm 0.77$ ppm short of what would be desired by experimental data.  Using the Simula {\it et al.} value for $\Delta_{\rm pol}$ would make the deficit greater.  Using other proton form factor fits (limitng ourselves to modern ones that fit the data well) in evaluating $\Delta_Z$ can reduce the deficit somewhat, but not by enough to ameliorate the problem~\cite{Nazaryan:2005zc}.

\begin{table}[htdp]
\caption{Corrections to hydrogenic hyperfine structure, as they could have been given in 2004.  The first line with numbers gives the ``target value'' based on the experimental data and the best evaluation of the Fermi energy (8 figures) based on known physical constants.   The corrections are listed next.  (The Zemach term includes a 1.53\% correction from higher order electronic contributions~\cite{Karshenboim:1996ew}, as well as a +0.07 ppm correction from muonic vacuum polarization and a +0.01 ppm correction from hadronic vacuum polarization~\cite{Volotka:2004zu}.)  The total of all corrections is $1.59 \pm 0.77$ ppm short of the experimental value. }
\begin{center}

\begin{tabular}{lrc}
Quantity						&	value (ppm)	& uncertainty (ppm) \\
\hline
$({E_{\rm hfs}(e^-p)}/{E_F^p})  - 1$	&	$1\ 103.49$  &	$0.01$	\\
\hline
$\Delta_{\rm QED}$				& $1\ 136.19$	&	$0.00$	\\
$\Delta_Z$ (using Friar \& Sick~\cite{Friar:2003zg})
							&	$-41.59$	&	$0.46$	\\
$\Delta_R^p$					&	$5.84$	&	$0.15$	\\
$\Delta_{\rm pol}$ (from Faustov \& Martyenko, 2002~\cite{Faustov:yp})		
							&	$1.40$	&	$0.60$	\\
$\Delta_{\rm weak}^p$			&	$0.06$	&	 		\\
\hline
Total							&  $1101.90$	&	$0.77$	\\
Deficit						&	$1.59$	&	$0.77$     \\
\hline
\end{tabular}

\end{center}
\label{table:two}
\end{table}%

The discrepancy is not large, measured in standard deviations.   On the other hand, the problem is clearly not in statistical fluctuations of the hfs measurement one is trying to explain, so one would like to do better.  As listed in the Table, the largest uncertainty in the corrections comes from $\Delta_{\rm pol}$.  Further, the polarizability corrections require knowledge of $g_1$ and $g_2$ at relatively low $Q^2$, and good data pressing farther into the required kinematic regime has relatively recently become available from JLab (the Thomas Jefferson National Accelerator Laboratory, in Newport News, Virginia, USA).  Accordingly, we shall present a state-of-the-art evaluation of the polarizability correction for electronic hydrogen.  To give away our results~\cite{Nazaryan:2005zc} at the outset, we essentially confirm (remarkably, given the improvements in data) the 2002 results of Faustov and Martynenko.

\section{Re-evaluation of $\Delta_{\rm pol}$}

\begin{figure}[htbp]
\begin{center}
\includegraphics[height=6cm]{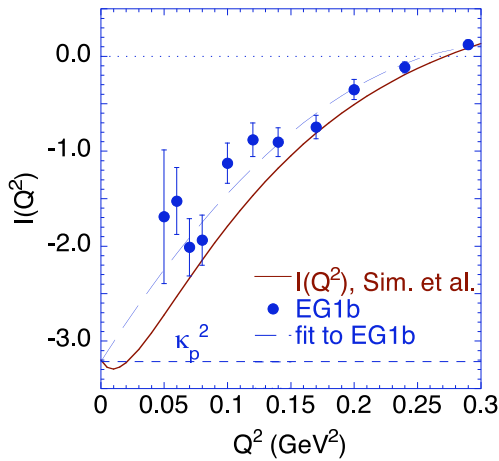}
\caption{Data for (an integral over) the spin-dependent struture function $g_1(\nu,q^2)$.  The data is from the EG1 experiment at Jefferson Lab and from the year 2005.  The Simula {\it et al.} curve is from a fit published in 2002.}
\label{fig:eg1}
\end{center}
\end{figure}

Data for $g_1(\nu,q^2)$ has improved due to the EG1 experiment at JLab,, which had a data run in 2000--2001.  
Some data based on preliminary analysis became available in 2005~\cite{deur}; final data is anticipated in late 2006.  A sample of the new data is shown in Fig.~\ref{fig:eg1}.  Since a function of two variables can be complicated to show, what is shown is the integral
\ba
I(Q^2) \equiv 4 m_p  \int_{\nu_{th}}^\infty \frac{d\nu}{\nu^2}  \,  g_1(\nu, -Q^2)  \,,
\ea
which differs from an integral appearing in $\Delta_1$ in lacking the auxiliary function.  The integration was done by the experimenters themselves.  We remind the reader that this integral is expected to reach the Gerasimov-Drell-Hearn value $-\kappa_p^2$ at the real photon point, and that because of cancellations the difference of this integral from $-\kappa_p^2$ is more relevant to the final answer for $\Delta_{\rm pol}$ than its absolute value.  

All the data shown is new;  there was no polarized electron-proton scattering data available below $Q^2 = 0.3$ GeV$^2$ when Simula {\it et al.} and Faustov and Martynenko did their earlier fits.  A curve obtained by integrating Simula {\it et al.}'s fit for $g_1$ is also shown on the Figure; we do not have enough information to produce a similar curve for Faustov and Martynenko's fit.

Integration in the region $Q^2 > 0.05$ GeV$^2$ is done using analytic fits to actual data for $g_1$.   For $Q^2$ below $0.05$ GeV$^2$, where there is no data, we do an interpolation based on a low $Q^2$ expansion within the integral to get (with $Q_1^2 \to 0.05$ GeV$^2$),
\ba
\Delta_1 [0,Q_1^2]  &\equiv& \frac{9}{4}\int_0^{Q_1^2} \frac{dQ^2}{Q^2}\left\{
	F_2^2(-Q^2) +4 m_p
		\int_{\nu_{th}}^\infty	 \frac{d\nu}{\nu^2} \beta(\tau)  g_1(\nu, -Q^2)
			\right\} 
					\\ 	\nonumber
&=&	  \left[ -\frac{3}{4} r_P^2 \kappa^2_p + 18 m_p^2 \, c_1
		- \frac{5m_p^2}{4 \alpha} \gamma_0 \right] Q_1^2 + {\cal O}(Q_1^4)  \,.
\ea
Here $r_P$ is from the expansion of the Pauli form factor
$
F_2(-Q^2) = \kappa_p^2 \left( 1 -  r_P^2 Q^2/6 + \ldots \right)
$,
and the ``forward spin polarizability'' $\gamma_0$ has been evaluated from data for other purposes~\cite{Drechsel:2002ar}, 
\ba
\gamma_0 \equiv   \frac{2\alpha}{m_p}    \int_{\nu_{th}}^\infty   \frac{d\nu}{\nu^4}  \  g_1(\nu,0)
	= \left[ -1.01 \pm 0.08 \pm 0.10 \right] \times 10^{-4} {\rm\ fm}^4  \,.
\ea
The parameter $c_1$ is defined from the slope at low $Q^2$ of the integral shown in Fig.~\ref{fig:eg1},
\ba
I(Q^2) = 4m_p \int_{\nu_{th}}^\infty   \frac{d\nu}{\nu^2} \  g_1(\nu,-Q^2)
	= - \kappa_p^2  + 8 m_p^2 \, c_1 Q^2 + {\cal O}(Q^4)  \,;
\ea 
we find and use $c_1 = 2.95 \pm 0.11$ GeV$^{-4}$~~\cite{Nazaryan:2005zc}. 

We need to comment that for $\Delta_2$, we need $g_2$, and there is almost no data for $g_2$ on the proton.  One estimates $g_2$ by relating it to $g_1$ using the Wandzura-Wilczek relation~\cite{Wandzura:1977qf}, which we shall not detail here.  Fortunately, the auxiliary function $\beta_2(\tau)$ is small over the region where we need to do the integrals, so that even when we assigned 100\% error bars to the contribution from $g_2$, the effect on the final answer was not great.

Our overall result is~\cite{Nazaryan:2005zc}
\ba
\Delta_{\rm pol} = 1.3 \pm 0.3 {\rm\ ppm} \,,
\ea
which is similar to the 2002 Faustov-Martynenko result.  This result means that the polarizability corrections no longer give the largest uncertainty in Table~\ref{table:two}.  It also
means that the theory deficit outlined in Table~\ref{table:two} still remains, even becoming modestly larger with a smaller uncertainty limit, at $(1.69 \pm 0.57)$ ppm.

Faustov, Gorbacheva, and Martynenko~\cite{Faustov:2006ve} quite recently published a new analysis and result for $\Delta_{\rm pol}$, obtaining the somewhat larger value
\ba
\Delta_{\rm pol} = 2.2 \pm 0.8 {\rm\ ppm} \,.
\ea
We still believe our published result~\cite{Nazaryan:2005zc} is the best one for now because the Jefferson Lab EG1b data, which goes to lower $Q^2$ than other data sets, has been used to constrain and validate the fits that we use to do the integrals.  Faustov {\it et al.} used only higher $Q^2$ data from other laboratories.  It is, of course, possible that the final EG1b data will lead to some change.

\section{Comments on the derivations of the formulas}


The polarizability corrections depend on theoretical results that are obtained using unsubtracted dispersion relations.  This has been alluded to before in this text, and this section will attempt to explain how a dispersion calculation works and what an unsubtracted dispersion relation is.  Also, given that there may be a small discrepancy between calculation and data, one would like to assess the validity of unsubtracted dispersion relations.  

Also, the hyperfine splitting in muonic hydrogen may be measured soon.
The polarizability corrections have been calculated for this case also~\cite{Faustov:2001pn}, albeit only with older fits to the structure function data and the relevant formulas, with non-zero lepton mass everywhere, are available~\cite{Faustov:2001pn} from a single source, so one would like to verify these formulas.  It turns out that keeping the lepton mass does not greatly increase calculational effort or the length of the formulas, so we can do the groundwork for the muonic hfs case simultaneously with the assessment of the ordinary hydrogen hfs calculation, although we shall not here display the formulas for non-zero lepton mass.

The calculation begins by writing out the loop calculation using the known electron vertices and the definition of the Compton scattering amplitudes involving $H_1$ and $H_2$ as given in Eq.~(\ref{eqn:ta}).  One can and should use this formalism for all the hadronic intermediate states, including the single proton intermediate states.  The single proton intermediate states give contributions to $H_1$ and $H_2$ that can be (more-or-less) easily calculated given a photon-proton-proton vertex such as Eq.~(\ref{vertex}).  For reference, we give the result for $H_1$,
\ba
\label{eq:el}
H_1^{el} = - \frac{2m_p}{\pi} \left( 
	\frac{ q^2 F_1(q^2) G_M(q^2) } { (q^2 + i\epsilon)^2 - 4m_p^2 \nu^2 }
		+ \frac{ F_2^2(q^2) }{ 4 m_p^2 }  \right)		\,.
\ea
The criticism of the proton vertex used to obtain the above result is that it is not demonstrably valid when the intermediate proton is off shell, so the above expression may or may not be correct overall.  However, it is correct at the proton pole.

One may do a unified calculation of the elastic and inelastic contributions.  Since we don't have a direct calculation of the $H_i$ for the inelastic case, we have to obtain them using dispersion relations.  Also obtaining the elastic terms from the dispersion relation is no problem~\cite{Iddings,Drell:1966kk}.  One just needs the imaginary parts of $H_i^{el}$;  these are easy to obtain, and contain Dirac delta-functions that ensure the elastic scattering condition $\nu = \pm Q^2/(2m_p)$ and hence depends only on the reliable part of Eq.~(\ref{eq:el}).

Dispersion relations involve imagining one of the real variables to be a complex one an then using the Cauchy integral formula to find the functions $H_i$ at a particular point in terms of an integral around the boundary of some region.  In the present case we ``disperse'' in $\nu$, treating $q^2$ as a constant while we do so.  Three things are needed to make the dispersion calculation work:
\begin{itemize}
\item The Cauchy formula and knowing the analytic structure of the desired amplitudes. 
\item The optical theorem, to relate the forward Compton ${\rm Im\,} H_i$ to inelastic scattering cross sections. 
\item Legitimately discarding contributions from some 
$\infty$ contour, if the dispersion relation is to be ``unsubtracted.''
\end{itemize}
The first two are not in question.

\begin{figure}[htbp]
\begin{center}
\includegraphics[height=5cm]{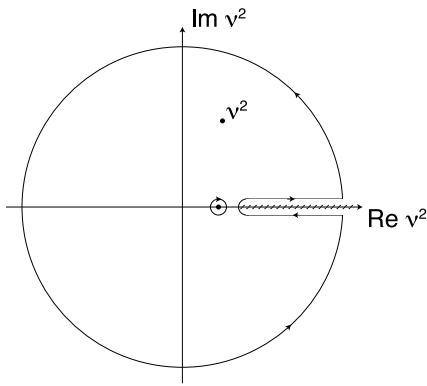}
\caption{Contour in complex $\nu^2$ plane for applying Cauchy identity to $H_1$ or $H_2$.}
\label{fig:cauchy}
\end{center}
\end{figure}

For the present case, the contour of integration is illustrated in Fig.~\ref{fig:cauchy},  where one should imagine the outside circle having infinite radius.   The result for $H_1$ begins its existence as
\ba
H_1(\nu,q^2) 
	&=& \frac{{\rm Res}  \left. H_1 (\nu,q^2) \right|_{el} }{\nu^2_{el} -\nu^2}
				\\[1.25ex]  \nonumber
&+& 
	  \frac{1}{\pi} \int_{cut} \frac{ {\rm Im\,} H_1(\nu',q^2)}{ {\nu'}^2-\nu^2} d{\nu'}^2
	+ \frac{1}{2\pi i}   \int_{|\nu'|=\infty}  \frac{ H_1(\nu',q^2) }{ {\nu'}^2-\nu^2 } d{\nu'}^2  \,.
\ea

The numerator of the first term is the residue (Res) from the poles in $\nu$ for the elastic part of $H_1$, as from Eq.~(\ref{eq:el}).  Note that the $F_2^2$ term in $H_1^{el}$, Eq.~(\ref{eq:el}), is constant in $\nu$, certainly lacking a pole in $\nu$.  Hence this term never enters the dispersion relation, and no $F_2^2$ term arises from the elastic contribution, as calculated this way.

The second term leads to the $g_1$ term in the quantity $\Delta_1$ given earlier, after using the optical theorem to relate ${\rm Im\,} H_i$ to $g_1$.

The third term is the integral over the part of the contour which is the infinite radius circle.  The commonly quoted results for $\Delta_{\rm pol}$, which appear in this talk, depend on dropping this term. The term is zero, if $H_1$ falls to zero at infinite $|\nu|$.   Assuming this is true, however, appears to be a dramatic assumption.  It fails for $H_1^{el}$ alone.  Hence, for the assumption to succeed requires an exact cancelation between elastic and inelastic contributions (or a failure of Eq.~(\ref{eq:el}) on the big contour).  On the positive side are several considerations.  One is that nearly the same derivation gives the GDH sum rule, which is checked experimentally and works, within current experimental uncertainty (8\%)~\cite{Pedroni:2006ta}. Also, the GDH sum rule is checkable theoretically in QED, where lowest order and next-to-lowest order perturbation theory calculations can be done, and the GDH sum rule appears to work~\cite{Dicus:2000cd,Altarelli:1972nc}.  Finally, Regge theory suggests the full Compton amplitude does fall to zero with energy~\cite{abarb67}, as one would like, although Regge theory famously gave wrong high $\nu$ behavior for spin-independent analogs of $g_1$ and $g_2$~\cite{Damashek:1969xj}.  Hence there are indications, though not decisive proof, supporting the unsubtracted dispersion relation.

The derivation finishes, as noted earlier, by subtracting a term involving $F_2^2$ from the relativistic recoil term, so as to obtain exactly the elastic corrections $\Delta_{el} = \Delta_Z + \Delta_R^p$ that were obtained (say) by Bodwin and Yennie for a calculation of the elastic terms only, using Eq.~(\ref{vertex}) at the photon-proton vertices and no dispersion theory~\cite{Bodwin:1987mj}. After adding the same term to the polarizability corrections in $\Delta_1$, one obtains the commonly quoted result for $\Delta_1$~\cite{Drell:1966kk,DeRafael:mc,Faustov:yp}.  Beyond the historical connection, if one is comfortable with the unsubtracted dispersion relation, the use of the dispersion theory gives a more secure result because it uses only the pole part of the photon-proton-proton vertex, so that the combined elastic and inelastic result does not depend on the general validity of whatever photon-proton-proton vertex one uses.


\section{Conclusion}


The evaluation of the polarizability contributions to hydrogen hyperfine structure, $\Delta_{\rm pol}$, based on latest proton structure function data is
$
\Delta_{\rm pol} = 1.3 \pm 0.3 {\rm\ ppm}
$~\cite{Nazaryan:2005zc}.
This is quite similar to the Faustov-Martyenko 2002 result,  
which we think is remarkable given the improvement in the data upon which it is based.  Most of the calculated $\Delta_{\rm pol}$ comes from integration regions where the photon four-momentum squared is small, $Q^2 < 1$ GeV$^2$.

There is still a modest discrepancy between the hydrogen hfs calculation and experiment, on the order of 2 ppm.  Optimistically, one can hope for a rapid reconciliation between data and calculation.  It surely has not been missed that using the Kelly form factor~\cite{sickandkelly} value of the Zemach radius and the new Faustov-Gorbacheva-Martynenko value for $\Delta_{\rm pol}$~\cite{Faustov:2006ve}
 give excellent agreement between theory and data.  Nonetheless, one can argue that other choices are currently better.  The integrals that give $\Delta_Z$ and $\Delta_{\rm pol}$ emphasize the low-$Q^2$ region.  The Sick form factors are the only modern ones that are tuned to fit best at low $Q^2$, and the determination of $\Delta_{\rm pol}$ in~\cite{Nazaryan:2005zc} is the only one that has explicitly used the lower $Q^2$ Jefferson Lab inelastic data.

An interplay between the fields of atomic and nuclear or particle physics may be relevant to sorting out problem.  For one example, the best values of the proton charge radius currently come from small corrections accurately measured in atomic Lamb shift~\cite{Mohr:2000ie}.  Sick's value of the charge radius~\cite{sickandkelly}, from the analysis of scattering data, is somewhat larger.  The precision of the atomic measurement of the proton charge radius can increase markedly if the Lamb shift is measured in muonic hydrogen~\cite{Antognini:2005fe}, which could happen in 2007, if the Paul Scherrer Institute accelerator schedule holds.  In the present context, the charge radius is noticed by its effect on determinations of the Zemach radius.

For ourselves, we look forward to a high accuracy resolution of the proton structure corrections to hydrogen hfs, and also to finishing a clear continuation of the  present program by the evaluation of the muonic hydrogen ground state hfs. We have formulas with all lepton masses in place, and are currently waiting until the final EG1 data is released, which we think will be rather soon, before proceeding and publishing a numerical evaluation.

My contributions to this subject have all been made in collaboration with Vahagn Nazaryan and Keith Griffioen.  I thank  them for the pleasure I have had working with them.  
In addition, we thank Jos\'e Goity,  Savely Karshenboim, Ingo Sick, Silvano Simula, and Marc Vanderhaeghen for
helpful discussions and information.
This work was supported by the National Science Foundation
under grants PHY-0245056 and PHY-0555600 (C.E.C.); PHY-0400332 (V.N.);
and by the Department of Energy under contract DE-FG02-96ER41003 (K.A.G.).

%


\begin{thebibliography}{8.}
\addcontentsline{toc}{section}{References}

\bibitem{Karshenboim:2005iy}
  S.~G.~Karshenboim,
  Phys.\ Rept.\  {\bf 422}, 1 (2005)
  [arXiv:hep-ph/0509010].

\bibitem{Volotka:2004zu}
A.~V.~Volotka, V.~M.~Shabaev, G.~Plunien and G.~Soff,
Eur.\ Phys.\ J.\ D {\bf 33}, 23 (2005).

\bibitem{dupays}
A.~Dupays, A.~Beswick, B.~Lepetit, C.~Rizzo, and D.~Bakalov,
Phys.\ Rev.\ A {\bf 68}, 052503 (2003).

\bibitem{Brodsky:2004ck}
  S.~J.~Brodsky, C.~E.~Carlson, J.~R.~Hiller and D.~S.~Hwang,
  Phys.\ Rev.\ Lett.\  {\bf 94}, 022001 (2005);
  Phys.\ Rev.\ Lett.\ {\bf 94}, 169902 (E) (2005)
  [arXiv:hep-ph/0408131].  See also~\cite{Friar:2005rs}.

\bibitem{Friar:2005rs}
  J.~L.~Friar and I.~Sick,
  Phys.\ Rev.\ Lett.\  {\bf 95}, 049101 (2005)
  [arXiv:nucl-th/0503020] and
  S.~J.~Brodsky, C.~E.~Carlson, J.~R.~Hiller and D.~S.~Hwang,
  Phys.\ Rev.\ Lett.\  {\bf 95}, 049102 (2005).
  
\bibitem{Eides:1995sq}
  M.~I.~Eides,
  Phys.\ Rev.\ A {\bf 53}, 2953 (1996).
  
\bibitem{Blunden:2003sp}
  P.~G.~Blunden, W.~Melnitchouk and J.~A.~Tjon,
  Phys.\ Rev.\ Lett.\  {\bf 91}, 142304 (2003)
  [arXiv:nucl-th/0306076];
 Y.~C.~Chen, A.~Afanasev, S.~J.~Brodsky, C.~E.~Carlson and M.~Vanderhaeghen,
  Phys.\ Rev.\ Lett.\  {\bf 93}, 122301 (2004)
  [arXiv:hep-ph/0403058];
 A.~V.~Afanasev, S.~J.~Brodsky, C.~E.~Carlson, Y.~C.~Chen and M.~Vanderhaeghen,
  Phys.\ Rev.\ D {\bf 72}, 013008 (2005)
  [arXiv:hep-ph/0502013];
 J.~Arrington,
  Phys.\ Rev.\ C {\bf 71}, 015202 (2005)
  [arXiv:hep-ph/0408261].
  
\bibitem{IddingsP}  C.~K.~Iddings and P.~M.~Platzman, Phys.\ Rev.\ {\bf 113}, 192 (1959).
  
\bibitem{Bodwin:1987mj}
G.~T.~Bodwin and D.~R.~Yennie,
Phys.\ Rev.\ D {\bf 37}, 498 (1988).

\bibitem{Martynenko:2004bt}
  A.~P.~Martynenko,
  Phys.\ Rev.\ A {\bf 71}, 022506 (2005)
  [arXiv:hep-ph/0409107].

\bibitem{Zemach} A.~C.~Zemach, Phys.\ Rev.\ {\bf 104}, 1771 (1956).

\bibitem{Karshenboim:1996ew}
S.~G.~Karshenboim,
Phys.\ Lett.\  {\bf 225A}, 97 (1997).

\bibitem{sickandkelly}
  I.~Sick,
  Phys.\ Lett.\ B {\bf 576}, 62 (2003)
  [arXiv:nucl-ex/0310008];
J.~J.~Kelly,
  Phys.\ Rev.\ C {\bf 70}, 068202 (2004).

\bibitem{Iddings} C.~K.~Iddings, Phys.\ Rev.\ {\bf 138}, B446 (1965).

\bibitem{Drell:1966kk}
S.~D.~Drell and J.~D.~Sullivan,
Phys.\ Rev.\  {\bf 154}, 1477 (1967).

\bibitem{DeRafael:mc}
E.~De Rafael,
Phys.\ Lett.\ B {\bf 37}, 201 (1971).

\bibitem{Gnaedig:qt}
P.~Gn\"adig and J.~Kuti,
Phys.\ Lett.\ B {\bf 42}, 241 (1972).

\bibitem{Faustov:yp}
R.~N.~Faustov and A.~P.~Martynenko,
Eur.\ Phys.\ J.\ C {\bf 24}, 281 (2002);
R.~N.~Faustov and A.~P.~Martynenko,
Phys.\ Atom.\ Nucl.\  {\bf 65}, 265 (2002)
[Yad.\ Fiz.\  {\bf 65}, 291 (2002)].

\bibitem{Nazaryan:2005zc}
  V.~Nazaryan, C.~E.~Carlson and K.~A.~Griffioen,
  Phys.\ Rev.\ Lett.\  {\bf 96}, 163001 (2006)
  [arXiv:hep-ph/0512108].

\bibitem{Anthony:2000fn}
  P.~L.~Anthony {\it et al.}  [E155 Collaboration],
  Phys.\ Lett.\ B {\bf 493}, 19 (2000)
  [arXiv:hep-ph/0007248].
  
\bibitem{Anthony:2002hy}
  P.~L.~Anthony {\it et al.}  [E155 Collaboration],
  Phys.\ Lett.\ B {\bf 553}, 18 (2003)
  [arXiv:hep-ex/0204028].

\bibitem{Fatemi:2003yh}
  R.~Fatemi {\it et al.}  [CLAS Collaboration],
  Phys.\ Rev.\ Lett.\  {\bf 91}, 222002 (2003)
  [arXiv:nucl-ex/0306019].

\bibitem{models}
J.\ Yun {\it et al.}, Phys.\ Rev.\ C {\bf 67} 055204, (2003); S. Kuhn, private communication.

\bibitem{deur}
A.~Deur, arXiv:nucl-ex/0507022.

\bibitem{Gerasimov:1965et}
  S.~B.~Gerasimov,
  Sov.\ J.\ Nucl.\ Phys.\  {\bf 2}, 430 (1966)
  [Yad.\ Fiz.\  {\bf 2}, 598 (1966)].

\bibitem{Drell:1966jv}
  S.~D.~Drell and A.~C.~Hearn,
  Phys.\ Rev.\ Lett.\  {\bf 16}, 908 (1966).
  
\bibitem{Simula:2001iy}
  S.~Simula, M.~Osipenko, G.~Ricco and M.~Taiuti,
  Phys.\ Rev.\ D {\bf 65}, 034017 (2002)
  [arXiv:hep-ph/0107036].  Silvano Simula provided
  us with an updated version of the code, including error estimates.

\bibitem{Friar:2003zg}
J.~L.~Friar and I.~Sick,
Phys.\ Lett.\ B {\bf 579}, 285 (2004).

\bibitem{Drechsel:2002ar}
  D.~Drechsel, B.~Pasquini and M.~Vanderhaeghen,
  Phys.\ Rept.\  {\bf 378}, 99 (2003)
  [arXiv:hep-ph/0212124].

\bibitem{Wandzura:1977qf}
  S.~Wandzura and F.~Wilczek,
  Phys.\ Lett.\ B {\bf 72}, 195 (1977).

\bibitem{Faustov:2006ve}
  R.~N.~Faustov, I.~V.~Gorbacheva and A.~P.~Martynenko,
   ``Proton polarizability effect in the hyperfine splitting of the hydrogen
  arXiv:hep-ph/0610332.
  
\bibitem{Faustov:2001pn}
  R.~N.~Faustov, E.~V.~Cherednikova and A.~P.~Martynenko,
  Nucl.\ Phys.\ A {\bf 703}, 365 (2002)
  [arXiv:hep-ph/0108044].
  
\bibitem{Pedroni:2006ta}
  P.~Pedroni  [GDH and A2 Collaborations],
  AIP Conf.\ Proc.\  {\bf 814}, 374 (2006);
 H.~Dutz {\it et al.}  [GDH Collaboration],
  Phys.\ Rev.\ Lett.\  {\bf 94}, 162001 (2005).

\bibitem{Dicus:2000cd}
  D.~A.~Dicus and R.~Vega,
  Phys.\ Lett.\ B {\bf 501}, 44 (2001)
  [arXiv:hep-ph/0011212];

\bibitem{Altarelli:1972nc}
  G.~Altarelli, N.~Cabibbo and L.~Maiani,
  Phys.\ Lett.\ B {\bf 40}, 415 (1972).
  
\bibitem{abarb67} 
  H. D. I. Abarbanel and S. Nussinov, Phys.\ Rev.\ {\bf 158}, 1462 (1967).

\bibitem{Damashek:1969xj}
  M.~Damashek and F.~J.~Gilman,
  Phys.\ Rev.\ D {\bf 1}, 1319 (1970);
  C.~A.~Dominguez, C.~Ferro Fontan and R.~Suaya,
  Phys.\ Lett.\ B {\bf 31}, 365 (1970).

\bibitem{Mohr:2000ie}
P.~J.~Mohr and B.~N.~Taylor,
Rev.\ Mod.\ Phys.\  {\bf 72}, 351 (2000);
and Rev.\ Mod.\ Phys.\ {\bf 77}, 1 (2005) [2002 CODATA]; {\it Cf.}, the electron scattering value of  I.~Sick,
  Phys.\ Lett.\ B {\bf 576}, 62 (2003)
  [arXiv:nucl-ex/0310008] 
  or the spread from Sick's value to that of
J.~J.~Kelly,
  Phys.\ Rev.\ C {\bf 70}, 068202 (2004).

\bibitem{Antognini:2005fe}
  A.~Antognini {\it et al.},
  AIP Conf.\ Proc.\  {\bf 796}, 253 (2005).



\end{thebibliography}
\end{document}